\documentclass[aps,prl,twocolumn,groupedaddress,amsmath,showpacs]{revtex4}
\usepackage{graphicx}
\begin{document}


\title{On the liquid-glass transition line in monatomic Lennard-Jones fluids}
\author{M. Robles}
\email{mrp@cie.unam.mx}
\author{M. L\'opez de Haro}
\email{malopez@servidor.unam.mx}
\affiliation{Centro de Investigaci\'on en Energ\'{\i}a\\
Universidad Nacional Aut\'onoma de M\'exico\\
AP 34, Temixco, Mor. CP 62580, M\'{e}xico.}
 
\date{\today}

\begin{abstract}
A thermodynamic approach to derive the liquid-glass transition line in the 
reduced 
temperature vs reduced density plane for a monatomic Lennard-Jones fluid is 
presented. The approach makes use of a recent reformulation of the classical
perturbation theory of liquids [M. Robles and M. L\'opez de Haro, Phys. Chem. 
Chem. Phys. {\bf 3}, 5528 (2001)] which is at grips with a rational function 
approximation for the Laplace transform of the radial distribution function 
of the hard-sphere fluid. The only input required is an equation of state for
the hard-sphere system. Within the Mansoori-Canfield/Rasaiah-Stell variational
perturbation theory, two choices for such an equation of state, leading to
a glass transition for the hard-sphere fluid, are considered. Good agreement
with the liquid-glass transition line derived from recent molecular dynamic 
simulations [Di Leonardo et al., Phys. Rev. Lett. {\bf 84}, 6054(2000)] is
obtained.
\end{abstract}


\maketitle


The liquid-glass transition in Lennard-Jones (LJ) fluids has been a subject of
interest for over twenty five years \cite{Rahman:1976}. Recently, Di
Leonardo et al. \cite{DiLeonardo:2000} using molecular dynamic simulations
determined the liquid-glass transition line of a monatomic LJ system in the
reduced density ($\rho^*$) vs. reduced temperature ($T^*$) plane. Here 
$\rho^*=\rho \sigma^3$ and $T^*=k_B T/\epsilon$, with $\rho$ the number 
density $k_{B}$ the Boltzmann constant, $T$ the temperature, and $\sigma$ 
and $\epsilon$ being the usual parameters of the LJ potential ($\phi_{LJ}(r)=4 
\epsilon (\sigma^{12}/r^{12} - \sigma^6/r^6)$, where $r$ is the distance).
Based on these results, they propose an off-equilibrium criterion to define 
the glass transition temperature $T_g$. This effective temperature compares 
rather  well with the $T_g$ obtained from equilibrium calculations.

In a completely different context, a long time ago Hudson and Andersen
\cite{Hudson:1978} addressed the nature of the glass transition in monatomic
liquids through an equilibrium calculation. In particular, in the case of the 
LJ fluid, they used the Weeks-Chandler-Andersen (WCA) perturbation theory of
liquids \cite{Weeks:1971} and two independent indications that a change in properties
similar to a glass transition happens in the hard-sphere (HS) fluid at a 
packing fraction of $0.533\pm0.014$. Given the then scarce amount of simulation
data available to compare with, their conclusion was that the use of the HS 
fluid as a reference fluid within the WCA scheme was adequate to derive the 
glass-transition line of monatomic fluids. As they already pointed out, for 
their approach to be useful in conection with glass formation in real systems,
it is crucial that the HS fluid itself undergoes a glass transition. As it is 
well known, the existence of a glass transition in the HS fluid has been a 
debatable issue for a long time, but evidence from various sources seems to 
suggest that this is indeed the case \cite{Speedy:1994,Baeyens:1997,Robles:1998,Speedy:1998,vanMegen:1993,vanMegen:1998}. 

The main aim of this letter is to assess whether an equilibrium calculation, 
such as the one carried out by Hudson and Andersen \cite{Hudson:1978}, can 
still be useful to describe the new data on the liquid-glass transition line 
of Lennard-Jones fluids. In a recent paper \cite{Robles:2001}, we have 
reformulated the most popular schemes of the perturbation theory of liquids, 
namely the Barker-Henderson \cite{Barker:1967}, the variational 
Mansoori-Canfiled/Rasaiah-Stell \cite{Mansoori:1969,Rasaiah:1970} and the 
WCA \cite{Weeks:1971} schemes, using the HS fluid as the 
reference fluid. Our study focussed on the equilibrium properties (equation 
of state, critical point, liquid and solid branches of the reduced temperature
vs. reduced density curve at coexistence and radial distribution function) of 
the Lennard-Jones fluid. All the calculations related to the different 
perturbative schemes that we reported in ref. \cite{Robles:2001}  rely on a 
known analytical rational function approximation (RFA) to the radial 
distribution function (rdf) $g_{HS}(r)$ for the HS fluid developed by Yuste 
and Santos \cite{Bravo:1991}. Their main idea is to propose that the Laplace 
transform $G(t)={{\mathcal L}}[r g_{HS}(r)]$, where $t$ is the Laplace transform 
variable, may be approximated by
\begin{equation}
G(t)=\frac{t}{12\eta}\frac{1}{1-e^t \Phi(t)},
\end{equation}
where $\eta=\pi/6\rho d^3$ is the packing fraction, $d$ being the hard-sphere 
diameter and $\Phi(t)$ is a rational function of the form
\begin{equation}
\Phi(t)=\frac{1+S_1 t+S_2 t^2+S_3 t^3 +S_4 t^4}{1+L_1 t+L_2 t^2}.
\label{phi}
\end{equation}
The coefficients $S_i$ and $L_i$ are algebraic functions of $\eta$. 
They are determined by imposing two physical restrictions to the rdf 
\cite{Robles:1997}, 
namely
\begin{enumerate}
\item The first integral moment of $h_{HS}(r)=g_{HS}(r)-1$ is well defined and
non-zero. 
\item The second integral moment of $h_{HS}(r)$ must guarantee the
thermodynamic
consistency of the compressibility factor $Z_{HS}=p/(\rho k_{B} T)$ (
$p$ being the pressure) and 
the isothermal susceptibility
$\chi_{HS}=(d(\eta Z_{HS})/d\eta)^{-1}$.
\end{enumerate}

These two conditions readily imply \cite{Bravo:1991,Robles:1997} that 

\begin{eqnarray}
\ L_{1} &=&\frac{1}{2}\frac{\eta +12\eta L_{2}+2-24\eta S_{4}}{2\eta +1},
\label{4a} \\
\ S_{1} &=&\frac{3}{2}\eta \frac{-1+4L_{2}-8S_{4}}{2\eta +1},  \label{4b} \\
S_{2} &=&-\frac{1}{2}\frac{-\eta +8\eta L_{2}+1-2L_{2}-24\eta S_{4}}{2\eta +1%
},  \label{4c} \\
S_{3} &=&\frac{1}{12}\frac{2\eta -\eta ^{2}+12\eta L_{2}(\eta-1)-1-72\eta ^{2}S_{4}}{\left( 2\eta +1\right) \eta },   \label{4d}\\
L_{2}&=&-3\left( Z_{HS}-1\right) S_{4},   \label{5}\\
S_{4}&=&\frac{1-\eta }{36\eta \left( Z_{HS}-1/3\right) }\times \nonumber \\ 
&&\left[ 1-\left[ 1+ \frac{Z_{HS}-1/3}{Z_{HS}-Z_{PY}}  \left( \frac{\chi }
{\chi _{PY}}-1\right) \right] ^{1/2}\right] .\label{6}
\end{eqnarray}
Here, $Z_{PY}=\frac{1+2\eta +3\eta ^{2}}{\left( 1-\eta \right) ^{2}}$ and $%
\chi _{PY}=\frac{\left( 1-\eta \right) ^{4}}{\left( 1+2\eta \right) ^{2}}$
denote the compressibility factor and isothermal susceptiblity arising in the
Percus-Yevick theory. In order to close the approximation, a given equation 
of state for a HS fluid {\it i.e.} an explicit expression for the 
compressibility factor $Z_{HS}$ is needed. 

All the perturbation schemes mentioned above introduce an effective (in general
density and temperature dependent) diameter of the spheres as a fitting 
parameter to adjust some of the thermodynamic properties of the system of 
interest with respect to those of the reference HS system. Once this effective
diameter is determined, one can infer by inversion the values of the 
temperature and density in the real system that correspond to a given packing 
fraction of the HS fluid. This fact was used  in ref. \cite{Robles:2001} to 
determine the liquid and solid branches of the reduced temperature vs. reduced 
density curve at coexistence for a LJ fluid from the knowledge of the 
packing fractions for the fluid-solid transition $\eta_{F-S}=0.494$ and the 
solid-fluid transition $\eta_{S-F}=0.54$ in the HS fluid, respectively 
\cite{Hansen:1969}. In a similar fashion, the liquid-glass transition line 
for the LJ system in the $\rho^*-T^*$ plane may be derived from the simple 
relationship
\begin{equation}
\frac{\pi}{6}\rho^* d ^{*3}(\rho^*,T^*)=\eta_g,
\label{cond1}
\end{equation}
where $\eta_g$ is the packing fraction corresponding to the glass transition in
the HS fluid and we have introduced the reduced diameter (in units of $\sigma$)denoted by $d^*$.

In order to proceed with such derivation, one has to specify the perturbation 
scheme and to know the value of $\eta_g$. As already stated above, Hudson and 
Andersen used the WCA and took $\eta_g=0.533\pm0.014$. In our case, we will 
consider two different equations of state for the HS fluid that allow us to 
calculate in a self-consistent way the value of $\eta_g$, and take the MC/RS 
scheme. 

The first such equation of state is the Pad\'e(4,3) constructed from the 
knowledge of the first eight virial coefficients \cite{vanRensburg:1993,
Sanchez:1994}. The compressibility factor corresponding to this equation of
state is given by

\begin{widetext} 
\begin{equation}
Z_{43}=\frac{1+1.024385 \eta+1.104537 \eta^2 -0.4611472 \eta^3 -0.7430382 \eta ^4}{1-2.985615\eta+3.00700\eta^2-1.097758\eta^3},
\end{equation}
\end{widetext}

Note that, as was pointed out in previous work\cite{Bravo:1996,Robles:1998}, 
within the RFA the Pad\'e(43) leads to a glass transition in the HS fluid at
$\eta_g=0.5604$. It should also be borne in mind that this equation of state
has a simple pole at a packing fraction very close to the fcc close-packing
fraction \cite{Sanchez:1994}.

On the other hand, the second equation of state is an {\it ad-hoc} approximation
constructed in the following way. Since the well known Carnahan-Starling(CS)
\cite{Carnahan:1969} 
equation of state has been shown to be very accurate throughout the complete 
fluid region and even in a small density range within the metastable regime 
but has a pole at $\eta=1$ which is clearly unphysical, we demand that
the new compressiblity factor $Z_{prop}$ has a pole at the random close 
packing fraction $\eta_{rcp}$ (up to this point taken as a parameter) and that 
the first eight coefficients in its series expaexpansionnsion coincide with the same 
coefficients in the series expansion of the Carnahan-Starling compressibility 
factor. Thus, $Z_{prop}(\eta)$ is taken to be of the form 
\begin{equation}
Z_{prop}(\eta)=\frac{1+\sum_{i=1}^{5}a_i \eta^i}{(1-\eta/\eta_{rcp})(1+b_1\eta+b_2 \eta^2)},
\label{eqprop}
\end{equation}
where the coefficients $a_i$ ($i=1,...,5$) and $b_j$ ($j=1,2$) depend on the
value of $\eta_{rcp}$. In order to determine this value, we
further assume that, according to the criterion that has been used 
earlier\cite{Bravo:1996,Robles:1998}, a glass transition in the HS fluid 
(considered to be a second order phase transition) occurs at $\eta_g$ 
(also unspecified at this stage) if $S4$ vanishes for $\eta_g$.  The above 
conditions lead to the following set of equations:
\begin{eqnarray}
S_4(\eta_g)&=&0, \label{condA} \\ 
Z_{glass}(\eta_g)&\equiv&\frac{A}{1-\eta_g/\eta_{rcp}} \nonumber \\ 
&=&Z_{prop}(\eta_g),\label{condB} \\ 
\chi_{glass}(\eta_g)&\equiv&\left(\frac{\partial( \eta Z_{glass})}{\partial \eta}\right)^{-1}\Big|_{\eta=\eta_g}\nonumber \\ 
&=&\frac{\chi_{PY}(\eta_g)(Z_{PY}(\eta_g)-1/3)}{(Z_{prop}(\eta_g)-1/3)},
\label{condC} 
\end{eqnarray} 
where $A$ is a constant and in writing the compressibility factor for the
glass we have taken the form proposed by Speedy \cite{Speedy:1994,Speedy:1998}.
The set of equations (\ref{condA})-(\ref{condC}) together with (\ref{eqprop}) 
allow  us to determine $\eta_{rcp}$, $A$ and $\eta_g$ in a self-consistent way.
The results are $\eta_{rcp}=0.6504$, $A=2.780$ and $\eta_g=0.5684$, which are
well within the range of published values for these quantities. In turn the
value of $\eta_{rcp}$ leads to the explicit form of $Z_{prop}(\eta)$, namely
\begin{widetext}
\begin{equation}
Z_{prop}(\eta)=\frac{1 + 0.153555\,\eta  - 0.428376\,{\eta }^2 - 
    2.7987\,{\eta }^3 - 0.317417\,{\eta }^4 - 
    0.105806\,{\eta }^5}{1 - 3.84644\,\eta  + 
    4.9574\,{\eta }^2 - 2.16386\,{\eta }^3}
\end{equation}
\end{widetext}
In the range $0\leq \eta \leq 1$ this equation of state only presents a simple 
pole at $\eta=\eta_{rcp}$. 

\begin{figure}[b]
\centerline{
  \includegraphics[height=5.1truecm,angle=0]{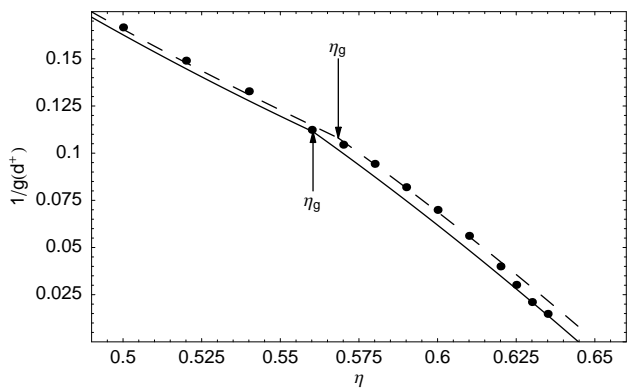}}
\caption{Inverse of the contact values of the rdf as derived from $Z_{43}$
(continuous line) and $Z_{prop}$(dashed line) as a function of $\eta$. The 
region corresponding to the glass has been obtained from $Z_{glass}$ and in 
the case of $Z_{43}$, $A=2.765$ and $\eta_{rcp}=0.6448$ \cite{Bravo:1996}.
The full symbols are simulation results of ref. \cite{Rintoul:1997} and the
arrows indicate the location of $\eta_g$ for each case.}
\label{fig1}
\end{figure}

In order to illustrate the numerical accuracy of these two equations of state
in the metastable region, in Fig. \ref{fig1} we display 
the inverse of the contact values of the rdf ($g(d^+)^{-1}=
4\eta/(Z_{HS}-1)$) derived from them as a function of the 
packing fraction and compare the results with the simulation data obtained by 
Rintoul and Torquato \cite{Rintoul:1997}. In this figure we have also included
the predictions of the corresponding equations of state for the glass adopting
the form suggested by Speedy\cite{Speedy:1994,Speedy:1998}.

For the sake of making the paper self contained, we now recall the condition
required to derive the temperature and density dependent diameter within the
MC/RS perturbation scheme (for details see refs. \cite{Mansoori:1969,
Rasaiah:1970,Robles:2001}). Here the effective diameter is 
chosen to minimize the Helmholtz free energy of the perturbed system.
To first order in $\beta=1/ T^*$ this diameter is determined from the 
equation \cite{Robles:2001}
\begin{multline}
\frac{\partial }{\partial d^*}\Big[ \int_{0}^{\rho^* }\frac{Z_{HS}-1}{\rho }
d \rho  + 4\left( \frac{1 }{d^*}\right) ^{6}\beta \times \\
\int_{0}^{\infty }
\frac{t}{1-e^{t}\Phi \left( t\right) }\left( \left( \frac{1
}{d^*}\right)^{6}\frac{t^{10}}{10!}-\frac{t^{4}}{4!}\right) dt \Big] =  0.
\label{mansoorieq}
\end{multline}
Clearly in solving this equation it should be understood that $Z_{HS}$ and
$\Phi(t)$ have been expressed as functions of $\rho^*$ and $d^*$.

\begin{figure}[t]
\centerline{
  \includegraphics[height=5.1truecm,angle=0]{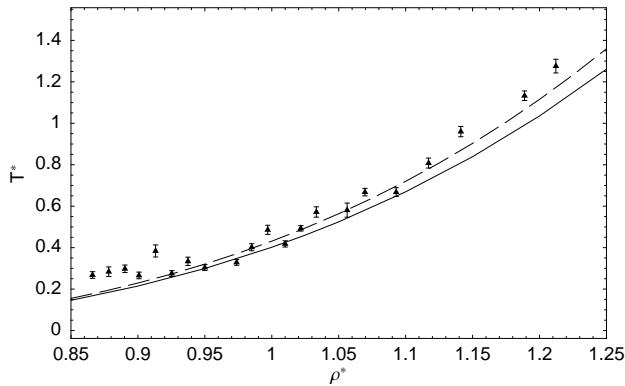}}
\caption{Liquid-glass transition lines as obtained with the MC/RS perturbation 
scheme using $Z_{43}$ (continuous line) and $Z_{prop}$(dashed line).  
The points are the simulation data of  Di Leonardo et al. in ref 
\cite{DiLeonardo:2000}}.
\label{fig2}
\end{figure}

Using the effective diameters determined from eq.(\ref{mansoorieq}) with 
either $Z_{43}$
or $Z_{prop}$ and the condition given by eq. (\ref{cond1}) (with the $\eta_g$ 
value corresponding to each equation of state) we have determined the
liquid-glass transition lines in the $\rho^*$-$T^*$ plane for the LJ fluid. 
These are
shown in Fig. \ref{fig2} where we have also included the recent simulation 
results of Di Leonardo et al. \cite{DiLeonardo:2000}. As clearly seen in the
figure, not only the qualitative trend observed in the simulations is 
reproduced, but also the quantitative agreement is rather good, specially for
the case of $Z_{prop}$. A remarkable aspect of these results is that they have
been derived self-consistently with no free parameters.

One question that immediately arises is to what extent the above results depend
on the MC/RS perturbation scheme and the choices for $Z_{HS}$. Concerning the
first issue, which is particularly relevant in view of the fact that the 
previous calculation by Hudson and Andersen \cite{Hudson:1978} used the WCA,
we have checked that the performance of this latter scheme is much poorer.
This may reflect that the choice of minimizing the Helmholtz free energy of
the perturbed system to derive the effective diameter may be more astringent 
than equating the isothermal compressiblities of the actual and the HS 
reference system. Something similar applies to the Barker-Henderson 
perturbation theory in which the lack of density dependence of the effective
diameter does not reproduce the desired trend. As far as the second issue is 
concerned, we have also performed calculations taking the CS equation of state
(which {\it does not} lead to a glass transition in the RFA formulation) and
adjusted the value of the packing fraction $\eta_g$ until good agreement with
the simulation data of Di Leonardo et al.\cite{DiLeonardo:2000} was obtained.
In fact for $\eta_g=0.55$ we get results pretty close to those derived using
$Z_{prop}$. The above suggests that if the exact equation of state
of the HS system (including a glass transition) were available then the use
of perturbation theory and the RFA approach would yield a very accurate 
description of the liquid-glass transition line of the LJ fluid. 

In conclusion, in this paper we have shown that an equilibrium approach to the
glass transition in LJ fluids, much in the same spirit as discussed by Baeyens 
and Verschelde \cite{Baeyens:1997} in the case of HS fluids, is wholly 
compatible with the molecular dynamics simulation results for the liquid-glass
transition line reported recently \cite{DiLeonardo:2000}. Perhaps the key
aspect of our approach is its self-consistency and the provision of a theory 
with no free parameters. Furthermore, the application of the same approach for
other monatomic fluids may be laborious, but in principle should follow the 
same steps.


\end{document}